\documentclass[twocolumn,showpacs,preprintnumbers,amsmath,amssymb,aps]{revtex4}

\usepackage{graphicx}
\usepackage{dcolumn}
\usepackage{bm}

\usepackage{hyperref}
\begin{document}


\title{Deterministic Controlled-NOT gate for single-photon two-qubit quantum
logic}

\author{Marco Fiorentino} \email{mfiore@mit.edu}
\author{Franco N.\ C.\ Wong}
\affiliation{Research Laboratory of Electronics, Massachusetts
Institute of Technology, Cambridge, MA 02139}

\begin{abstract}
We demonstrate a robust implementation of a deterministic
linear-optical Controlled-NOT (CNOT) gate for single-photon
two-qubit quantum logic.  A polarization Sagnac interferometer
with an embedded 45$^{\circ}$-oriented dove prism is used to
enable the polarization control qubit to act on the momentum
(spatial) target qubit of the same photon.  The CNOT gate requires
no active stabilization because the two spatial modes share a
common path, and it is used to entangle the polarization and
momentum qubits.
\end{abstract}


\maketitle

Knill, Laflamme, and Milburn \cite{KLM} show that probabilistic
two-qubit operations implemented in linear optical circuits with
ancilla photons can be used to build a scalable quantum computer.
Their work has stimulated much attention on the experimental
realization of linear optics quantum computation (LOQC) protocols,
and simple two-qubit gates based on quantum interference and
postselection have been demonstrated \cite{Franson}.

It has been recognized that building a deterministic QIP is
possible for systems that use several degrees of freedom of a
single photon to encode multiple qubits \cite{detth}.  This type
of QIP cannot be used for general-purpose quantum computation
because it requires resources that grow exponentially with the
number of qubits. Nevertheless, there is a growing interest in
few-qubit QIPs.  Small deterministic QIPs have been used to
demonstrate basic quantum logic protocols \cite{detex}. Several
authors have suggested the use of three- or four-qubit QIP based
on two degrees of freedom of a photon in combination with sources
of hyperentangled photons (i.e., photons that are entangled in
more than one degree of freedom) for an all-or-nothing
demonstration of nonlocality \cite{Zeilinger}, complete Bell's
state measurements \cite{Monken}, and quantum cryptography
\cite{genovese}. Small QIPs have also been proposed as a way to
implement quantum games \cite{HPgame}. Kim \cite{kim} has
demonstrated the creation of entanglement between the polarization
and momentum qubits of a single photon by introducing two possible
paths for the single photon and allowing the phase and
polarization for each path to be controlled separately.  This
method may have limited application potential due to the
interferometric arrangement of the two paths.

The utility of single-photon two-qubit (SPTQ) QIP depends on one's
ability to create and manipulate the individual qubits (qubit
rotation) and to implement two-qubit quantum gates in a robust
experimental setup.  In this Letter, we demonstrate the first
experimental realization of a deterministic linear optical
Controlled-NOT (CNOT) gate for two-qubit photons, with the
polarization serving as the control qubit and the momentum as the
target qubit.  Our implementation eliminates a roadblock for SPTQ
quantum logic protocols by using a simple and compact polarization
Sagnac interferometer with an internal dove prism oriented at
45$^{\circ}$ that requires no stabilization of the path lengths.
This SPTQ CNOT gate can be utilized to perform a variety of
quantum logic operations that are necessary for SPTQ QIP
implementation.  In particular, we have used it to demonstrate the
ease and efficiency of entangling the polarization and momentum
qubits of a single photon.  Finally, the SPTQ CNOT gate is
especially suitable for use with entangled photon pairs for
creating more complex entangled states of three or more qubits and
opens the way to the implementation of SPTQ procotols for
few-qubit QIP.

\begin{figure}[ht]
\scalebox{.35}{\includegraphics{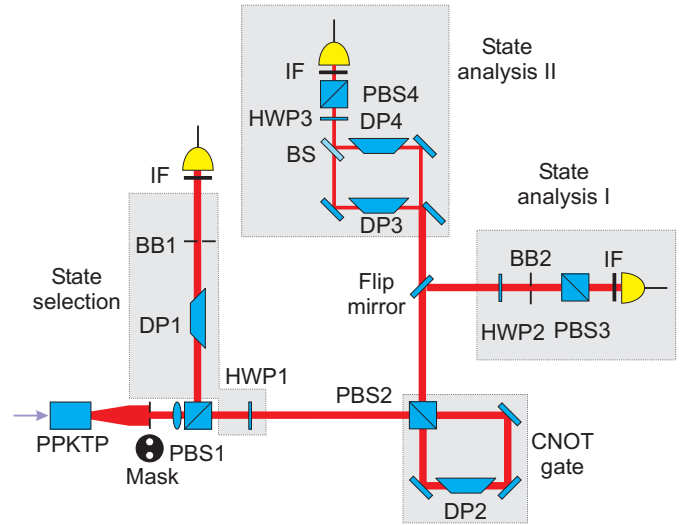}} \caption{Schematic of
experimental setup. PPKTP: periodically poled KTP crystal. PBS:
polarizing beam splitter. HWP: half-wave plate. DP: dove prism.
BS: 50-50 beam splitter. BB: beam block that allows one to block
the $L$ or $R$ part of the beam. IF: 1-nm interference filter
centered at 797 nm.} \label{Setup}
\end{figure}

As shown in Fig.~\ref{Setup}, the CNOT gate is a polarization
Sagnac interferometer containing a dove prism whose base is
oriented at a 45$^{\circ}$ angle relative to the horizontal plane.
This interferometer is similar to a non-polarizing Sagnac
interferometer for measuring the spatial Wigner function
\cite{ring}.  The input polarizing beam splitter (PBS2) directs
horizontally (vertically) polarized input light to travel in a
clockwise (counterclockwise) direction. As viewed by each beam,
the dove prism orientation is different for the two
counter-propagating beams such that the transformation of the
input spatial image differs for the horizontal ($H$) and vertical
($V$) polarizations.  Specifically, the top-bottom ($T$--$B$)
sections of the input beam are mapped onto the right-left
($R$--$L$) sections of the output beam for $H$-polarized light but
onto the $L$-$R$ sections for $V$-polarized light.  If we identify
$|H\rangle$, $|T\rangle$, and $|R\rangle$ with the logical
$|0\rangle$ and $|V\rangle$, $|B\rangle$, and $|L\rangle$ with the
logical $|1\rangle$, this setup implements a
polarization-controlled NOT (P-CNOT) gate in which the
polarization is the control qubit and the momentum (or spatial)
mode is the target qubit.  A key advantage of the Sagnac
interferometer is that the same amount of phase delay is added to
the two counterpropagating beams, thus removing the need for
active control of the interferometer.  We should point out that
the complementary momentum-controlled NOT (M-CNOT) gate in which
the momentum is the control qubit and the polarization is the
target qubit can be realized with a half-wave plate (HWP) oriented
at 45$^\circ$ relative to the horizontal position and inserted in
the path of the $B$ beam. It can also be shown \cite{next} that it
is possible, using linear optical components, to perform arbitrary
qubit rotation on either the polarization or momentum qubit
without disturbing the state of the other qubit of a single
photon.

Figure~1 shows the experimental setup for demonstrating the CNOT
gate operation. A 10-mm-long periodically poled hydrothermally
grown potassium titanyl phosphate (KTP) crystal (9.0-$\mu$m
grating period) was continuous-wave (cw) pumped at 398.5 nm for
type-II collinear frequency-degenerate parametric down-conversion
\cite{twos}.  We have previously obtained a signal-idler quantum
interference visibility of 99\% for this source.  The
idler-triggered signal beam was used as a single-photon source for
the input to the CNOT gate. The orthogonally polarized signal and
idler photons were separated by PBS1 in Fig.~1, and the signal
polarization could be rotated by HWP1.  We used postselection to
set the signal momentum ($T$ or $B$). We first had to verify that
the down-converted photon pairs show correlation in momentum
($T$--$B$). To do this we collimated the down-converted photons
and sent them through a mask with two apertures (1.5 mm diameter,
2.5 mm center-to-center distance) that defined the top and bottom
beams. After the PBS1 the $H$-polarized signal and $V$-polarized
idler were separated, spectrally filtered with a 1-nm interference
filter (IF) centered at 797\,nm, and detected by two Si
photon-counting modules (Perkin Elmer SPCM-AQR-14). Signal-idler
coincidence counts were measured with a fast AND gate based on
positive emitter coupled logic (PECL) with a coincidence window of
$\sim$1\,ns \cite{Taehyun}. The 1-ns coincidence window and
singles count rates of less than $10^5$/s ensure negligible
accidental coincidence counts.  When we blocked the top (bottom)
idler beam after PBS1 to pass the bottom (top) idler photons, only
the top (bottom) signal beam yielded coincidences, thus
establishing the momentum correlation between the signal and idler
photons. We used the same collimation and mask setup for the input
to the CNOT gate.  By blocking either the $T$ or $B$ section of
the idler with the beam blocker (BB1), we were able to postselect
the momentum state of the signal photon to be $T$ or $B$,
respectively.

Momentum entanglement \cite{ratap} in spontaneous parametric
down-conversion is the basis for the generation of polarization
entanglement from a noncollinearly phase-matched down-conversion
crystal by use of overlapping cones of emitted photons
\cite{kwiat}.  One can therefore take advantage of the natural
momentum entanglement of down-converted photon pairs to generate
more complex entangled states.  The state of a pair of
down-converted photons is:
\begin{eqnarray}
|\Theta\rangle = \frac{1}{\sqrt{2}} \left(|0110\rangle +
|0011\rangle\right)\,,
\end{eqnarray}
where the four qubits are written in the form of $|P_S M_S P_I M_I
\rangle$ for the  polarization ($P$) and momentum ($M$) qubits of
the signal ($S$) and idler ($I$) photons.  Applying SPTQ quantum
logic to a pair of entangled photons presents some interesting
possibilities.  For example, a swap operation can be realized with
a sequence of three CNOT gates.  If we apply the swap operation (a
P-CNOT gate, followed by a M-CNOT gate and another P-CNOT gate) to
both photons of $|\Theta\rangle$, we transform the
momentum-entangled state to a polarization-entangled state
\begin{eqnarray}
|\Theta\rangle_{Swap} = \frac{1}{\sqrt{2}} \left(|1001\rangle +
|0011\rangle\right).
\end{eqnarray}
Another example is the application of a single M-CNOT gate to both
signal and idler photons to obtain the four-qubit
Greenberger-Horne-Zeilinger (GHZ) state
\begin{eqnarray}
|\Theta\rangle_{GHZ} = \frac{1}{\sqrt{2}} \left(|1110\rangle +
|0001\rangle\right).
\end{eqnarray}

We characterized the CNOT gate by injecting in turn each of the
four natural basis states of the SPTQ $|PM\rangle$ basis and
analyzing the output state by means of BB2 and HWP2. Single-photon
input is guaranteed by signal-idler coincidence counting. We note
that for the PBS used in the CNOT gate setup, the $H$-polarized
light suffered a $\sim$10\% transmission loss ($\sim$5\% per
passage through PBS2). This imbalance was corrected by inserting
inside the Sagnac interferometer a thin glass plate near the
Brewster angle for the horizontal polarization and rotating it to
introduce an appropriate amount of differential loss between the
two polarizations. Figure~2 displays the measured truth table of
the CNOT gate, showing clearly the expected behavior of the gate.
For each input state the probability sum of the erroneous outcomes
is only $\sim$1\%. There is a slight asymmetry between the $H$-
and $V$-polarized outputs because the thin-plate compensation was
not perfect. The table of truth we measured is a necessary
consequence but not a sufficient proof that the device is a CNOT
gate, whose complete characterization requires quantum process
tomography \cite{QPT} that is experimentally challenging and time
consuming. However, as an indirect proof of its CNOT
functionality, it is adequate to demonstrate that the gate
preserves quantum coherence and creates entanglement.

\begin{figure}
\rotatebox{270}{\scalebox{.35}{\includegraphics{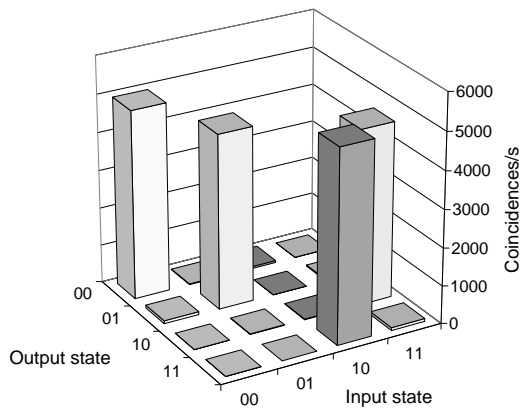}}}
\caption{Coincidence count rates as a function of the projected
output state for a given input state.} \label{Matrix}
\end{figure}

\begin{figure}
\rotatebox{270}{\scalebox{.35}{\includegraphics{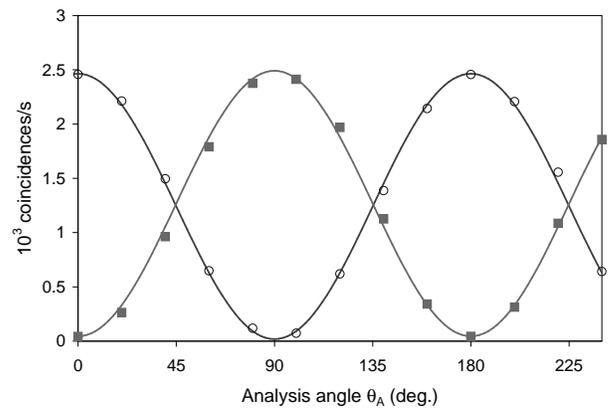}}}
\caption{Characterization of the entangled state
$|\Psi\rangle_{OUT}$ by projecting it onto $|\Psi\rangle_{1}$
(circles) and $|\Psi\rangle_{2}$ (squares) versus analysis angle
$\theta_A$. The lines are fits to the data.} \label{Entanglement1}
\end{figure}

It is well known that a CNOT gate can be used to entangle two
qubits.  Indeed for an input product state
\begin{equation}
|\Psi\rangle_{IN} = \frac{1}{\sqrt{2}}(|0\rangle_C + |1\rangle_C)
\otimes |0\rangle_T\,,
\end{equation}
where the subscripts $C$ and $T$ refer to the control
(polarization) and target (momentum) qubits, respectively, the
CNOT-gate output is the entangled state
\begin{equation}
|\Psi\rangle_{OUT} = \frac{1}{\sqrt{2}}(|0\rangle_C |0\rangle_T +
|1\rangle_C |1\rangle_T)\,.
\end{equation}
To create the input state $|\Psi\rangle_{IN}$ HWP1 was set to
rotate the signal polarization by 45$^{\circ}$ and the $T$-path of
the idler beam was blocked, thus generating the desired state
$|\Psi\rangle_{IN}$ by post selection. For the output state
analysis, we used BB2, HWP2, and PBS3 (state analysis I in Fig.~1)
to project the output onto
\begin{eqnarray}
|\Psi\rangle_1 &=& [\cos\theta_A|0\rangle_C +
\sin\theta_A |1\rangle_C] \otimes |0\rangle_T\,, \\
|\Psi\rangle_2 &=& [\cos\theta_A|0\rangle_C + \sin\theta_A
|1\rangle_C] \otimes |1\rangle_T\,,
\end{eqnarray}
where $\theta_A$ is polarization analysis angle (equal to twice
the angle setting) of HWP2.  We measured the projected output as a
function of $\theta_A$ and obtained the expected curves of the
signal-idler coincidence counts shown in Fig.~\ref{Entanglement1}.
The fits show a visibility of $98 \pm 1 \%$ for $|\Psi\rangle_{1}$
and $96.2 \pm 0.8 \%$ for $|\Psi\rangle_{2}$. These measurements
alone, however, do not imply that the output state is entangled,
as a mixture of the two states $|0\rangle_C |0\rangle_T$ and
$|1\rangle_C |1\rangle_T)$ would yield the same projective
results.

\begin{figure}
\rotatebox{270}{\scalebox{.35}{\includegraphics{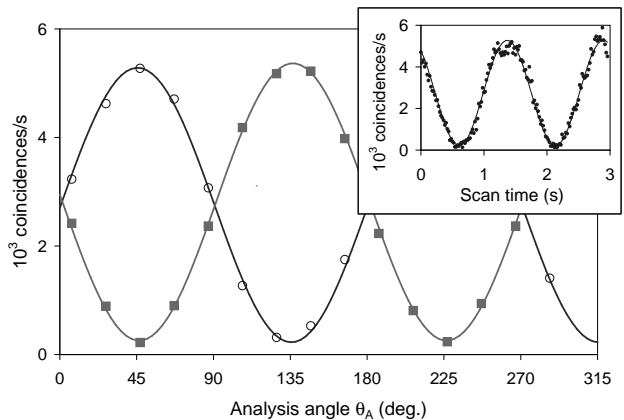}}}
\caption{Characterization of the entangled state
$|\Psi\rangle_{OUT}$ by projecting it onto $|\Psi\rangle_{3}$
(circles) and $|\Psi\rangle_{4}$ (squares) for different analysis
angles $\theta_A$. The lines are fits to the data. A pump power
higher than that used for Fig.~\ref{Entanglement1} accounts for
the higher coincidence counts. Inset: raw data for an
interferometer scan with $\theta_A = 25^ \circ$; the line is a fit
to the data.} \label{Entanglement2}
\end{figure}

Similar to a test of two-photon polarization entanglement, it is
necessary to make a projective measurement along $|0\rangle \pm
|1\rangle$ in the momentum space with states such as
\begin{eqnarray}
|\Psi\rangle_3 &=& [\cos\theta_A|0\rangle_C + \sin\theta_A
|1\rangle_C] \otimes \frac{(|0\rangle_T + |1\rangle_T)}
{\sqrt{2}}\,, \\
|\Psi\rangle_4 &=& [\cos\theta_A|0\rangle_C +
\sin\theta_A |1\rangle_C] \otimes \frac{(|0\rangle_T -
|1\rangle_T)}{\sqrt{2}}
\end{eqnarray}
and verify that the coincidences follow the expected curves. To
project the output state $|\Psi\rangle_{OUT}$ onto
$|\Psi\rangle_{3}$ and $|\Psi\rangle_{4}$ we used the state
analysis setup II shown in Fig.~\ref{Setup}.   This was an
interferometer for overlapping the $L$ and $R$ output beams of the
CNOT gate at a 50-50 beam splitter. The two arms of the
interferometer must be matched to within a coherence length of the
downconverted photons, which, in our case, was determined by the
1-nm IF placed in front of the detector. Two dove prisms, one that
flipped the beam vertically and the other horizontally, were
placed in the interferometer arms to obtain the proper orientation
for the spatial overlap of the CNOT $L$ and $R$ outputs. By
scanning the length of one arm of the interferometer and passing
the photons through HWP3 and PBS4 the state was projected onto
\begin{equation}
|\Psi\rangle_{\phi} = [\cos\theta_A|0\rangle_C + \sin\theta_A
|1\rangle_C] \otimes \frac{(|0\rangle_T + e^{i \phi}
|1\rangle_T)}{\sqrt{2}}\,,
\end{equation}
where the phase $\phi$ is determined by the path length difference
between the two arms of the interferometer and $\theta_A$ is the
polarization analysis angle.  For $\phi= 0$ and $\phi=\pi$ we
obtain states $|\Psi\rangle_{3}$ and$|\Psi\rangle_{4}$,
respectively. In the measurements we recorded the detected
coincidences while scanning the length of one interferometer arm
for different settings of the polarization analyzer (see inset of
Fig.~\ref{Entanglement2} for one of such traces). For each
$\theta_A$ the data points are fitted with the curve $\alpha+\beta
\sin ^2(\delta t + \gamma)$ where $\alpha$, $\beta$, $\delta$, and
$\gamma$ are fit parameters and $t$ is the scan time. The fitted
maxima and minima are plotted in Fig.~\ref{Entanglement2}. The
fitted curves in Fig.~\ref{Entanglement2} yield a visibility of
$90.8 \pm 0.8 \%$ for $|\Psi\rangle_{3}$ and $91.7 \pm 1.6 \%$ for
$|\Psi\rangle_{4}$, indicating an excellent entangling capability
of the CNOT gate. The curves are shifted by about $1^\circ$ with
respect to the expected value of $45^\circ$. We attribute this
shift to an imbalance in the interferometer due to a deviation of
the beam splitter reflectivity from 50\%.

In conclusion we have presented an implementation of a
deterministic CNOT gate that uses the polarization and momentum
degrees of freedom of a single photon as the control and target
qubits, respectively.  Based on a polarization Sagnac
interferometer with an embedded dove prism, the gate is built with
linear optical components, is robust and does not require any
active stabilization.  This device, in conjunction with sources of
entangled photons, can be utilized to create the four-qubit GHZ
state, to investigate nonstatistical violation of Bell's
inequality \cite{Zeilinger}, and for efficient two-photon Bell's
state analysis \cite{Monken}. We believe these more complex
entanglement sources together with single-photon two-qubit quantum
logic can be used to perform special computational tasks in a
variety of few-qubits quantum computing and quantum communication
protocols.

The authors would like to acknowledge K. Banaszek and V.
Giovannetti for useful discussion.  This work was supported by the
DoD Multidisciplinary University Research Initiative (MURI)
program administered by the Army Research Office under Grant
DAAD-19-00-1-0177.

\end{document}